# Effect of pressure on octahedral distortions in *R*CrO$_3$ (*R* = Lu, Tb, Gd, Eu, Sm): The role of *R*-ion size and its implications

Venkata Srinu Bhadram[1], Diptikanta Swain[1†], R. Dhanya[1†], Maurizio Polentarutti[2], A. Sundaresan[1], Chandrabhas Narayana[1*]

[1]*Chemistry and Physics of Materials Unit, Jawaharlal Nehru Centre for Advanced Scientific Research, Jakkur P.O., Bangalore 560064 India*

[2]*Elettra - Sincrotrone Trieste S.C.P.A., Strada Statale 14 - km 163,5 in AREA Science Park 34149 Basovizza, Trieste, Italy.*

† These authors contributed equally.

Email: cbhas@jncasr.ac.in

**Abstract:**

The effect of rare-earth ion size on the octahedral distortions in rare-earth chromites (*R*CrO$_3$, *R* = Lu, Tb, Gd, Eu, Sm) crystallizing in the orthorhombic structure has been studied using Raman scattering and synchrotron powder x-ray diffraction up to 20 GPa. From our studies on *R*CrO$_3$ we found that the octahedral tilts (distortions) increase with pressure. This is contrary to the earlier report which suggests that in LaCrO$_3$, the distortions decrease with pressure leading to a more ordered phase at high pressure. Here we observe that the rate of increase in distortion decreases with the increase in *R*-ion radii. This occurs due to the reduction in the compression of *R*O$_{12}$ polyhedra with a corresponding increase in the compression of the CrO$_6$ octahedra with increasing *R*-ion radii. From the Raman studies, we predict a critical *R*-ion radii, above which we expect the distortions in *R*CrO$_3$ to reduce with increasing pressure leading to what is observed in the case of LaCrO$_3$. These Raman results are consistent with our pressure dependent structural studies on *R*CrO$_3$ (*R* = Gd, Eu, Sm). Also, our results suggest that the pressure dependence of Néel temperature, $T_N^{Cr}$, (where the $Cr^{3+}$ spin orders) in *R*CrO$_3$ is mostly affected by the compressions of Cr-O bonds rather than the alteration of octahedral tilts.

## 1. Introduction:

Perovskites with general formula $ABO_3$ are fascinating class of multifunctional materials as they exhibit a wide variety of physical properties, e.g. magnetism, ferroelectricity, multiferroicity and superconductivity.[1,2] These properties are sensitive to the perovskite structure and $BO_6$ octahedral tilting distortions which can be controlled by external pressure, temperature and variations in A and B cations.[3,4] Among perovskites, rare-earth based compounds ($RMO_3$; $R$ is rare-earth, M is transition metal) have attracted lots of attention recently due to their multifunctional properties which have fascinating device applications.[5-7] Especially, $RMnO_3$ exhibit multiferroicity which depends on the size of the rare-earth. For example, $YMnO_3$ which crystallizes in hexagonal ($P6_3cm$) symmetry exhibits geometric ordering driven multiferroicity[8] whereas, magnetic ordering driven multiferroicity is observed in orthorhombic (*Pnma*), e.g. $TbMnO_3$.[9] Interestingly, external pressure can modify the octahedral distortions and can turn the hexagonal manganites into orthorhombic ones.[10] In other words, pressure as a parameter can alter the multiferroic properties of these systems.

Rare-earth orthochromites $RCrO_3$ ~~(*R* rare earth)~~ which are isostructural to orthorhombic manganites have a wide variety of applications in catalysis,[11] fuel cells,[12] etc. Recently, these materials were found to be multiferroic below the Néel temperature, $T_N^{Cr}$, (where the $Cr^{3+}$ spin orders) only when rare-earth is magnetic ion (i.e Gd, Sm).[13] It was proposed that the polarization induced at $T_N^{Cr}$ is possibly due to the exchange interactions between the magnetic $R^{3+}$ ions and the ordered $Cr^{3+}$ sublattice mediated through spin-phonon coupling.[14] Moreover, the spin configuration in $RCrO_3$ can also play a crucial role in inducing ferroelectricity.[13] It is thus logical to study the structural changes in $RCrO_3$ either by doping (chemical pressure) or by applying hydrostatic pressure. External pressure could indeed be considered as a fine tuning of the chemical substitution, hence would give information for both the doping as well as the volume effects.

In the past, many studies[15-19] have been aimed at understanding the pressure behaviour of orthorhombic perovskites and a general rule has also been proposed.[17] It predicts that the octahedral tilting distortions decrease with pressure in the case of trivalent (3:3 group) perovskites (e.g. $GdAlO_3$). A comparative study [19] of pressure evolution of octahedral distortions in $YMO_3$ (M=Al, Cr, Ti) negates this generalization and suggests the importance of size of B-cation radii for predicting the pressure induced structural changes in orthorhombic $ABO_3$ perovskites. In the same report, it is shown that structure of $YCrO_3$ becomes more distorted at higher pressures. On the contrary, a recent report[20] on $LaCrO_3$ suggests that it undergoes structural transformation to a higher symmetry phase (rhombohedral, $R\bar{3}c$) at around 5 GPa indicating the importance of *R*-ion radii in understanding the high pressure behaviour of $RCrO_3$. However, there were no studies reported on the role of *R*-ion radii in the pressure dependent orthorhombic distortion despite its obvious relation with the degree of octahedral tilts.[18]

Furthermore, $Cr^{3+}$-O-$Cr^{3+}$ exchange interactions are sensitive to the octahedral tilt angles[21] and thereby play a crucial role in the pressure induced changes in the magnetic transition temperature ($T_N^{Cr}$), which would regulate the magnetic properties of these materials. Although there are a good amount of work done in this line in the case of rare-earth nickelates ($RNiO_3$),[22] ferrites ($RFeO_3$),[23] and manganites ($RMnO_3$),[24] no such studies exist in the case of $RCrO_3$ to the best of our knowledge. Thus, the aim of the present studies is of two folds: i) to see the effect of rare-earth size on the pressure induced changes in the octahedral tilts ii) to be able to correlate these changes in the octahedral tilts to the magnetic properties of $RCrO_3$ in the temperature and pressure space.

In our study, we have used pressure dependent Raman mode behaviour of various orthochromites ($RCrO_3$) up to 20 GPa. Raman scattering is very sensitive to the structural distortions and is a unique tool for pressure dependent structural studies on perovskites. Also, we have used synchrotron x-ray diffraction data of $RCrO_3$ (*R*- Gd, Eu and Sm) to corroborate our results based on Raman studies. We have carefully selected the compounds; $LuCrO_3$, $TbCrO_3$, $GdCrO_3$, $EuCrO_3$ and $SmCrO_3$ for our present Raman studies. The selection, here, is entirely based on the rare earth ion size ($r_{R^{3+}}$: $Lu^{3+} < Tb^{3+} < Gd^{3+} < Eu^{3+} < Sm^{3+}$).

## 2. Experimental Details:

Powder samples of all the rare-earth chromites studied here were prepared using solid state method. The details of the sample synthesis are given in Ref.13. The pellets of these powder samples were crushed and pieces of required size were selected for high pressure studies. A membrane type high pressure diamond anvil cell (DAC) from BETSA, France has been used for the high pressure experiments. The diamond culet size is 400 microns. Methanol, ethanol (4:1) mixture has been used as a pressure transmitting medium and for *in situ* pressure measurements a ruby chip has been loaded along with sample inside DAC. Unpolarized Raman spectra were recorded in the back scattering geometry using a custom-built Raman spectrometer [25] equipped with a laser with excitation wavelength 532 nm and laser power of

~ 5 mW outside the DAC focused using a Nikon L Plan super long working distance 50x objective with a working distance of 17 mm and a numerical aperture of 0.45. It should be noted that the power density inside the DAC would be much smaller. The Raman spectra were deconvoluted using multiple Lorentzian functions with a proper background correction. Structural characterization under pressure was carried out using monochromatic ($\lambda$=0.7Å) high pressure x-ray powder diffraction beamline at Elettra synchrotron, Trieste. The pressure medium used here was silicone oil. Gold powder has been used for *in situ* pressure determination. Diffraction patterns were recorded on image plate and the collected patterns were integrated using FIT2D[26] program to get intensity vs 2θ plots.

## 3. Results and discussion:

### *3.1. Raman scattering studies:*

Rare-earth orthochromites ($RCrO_3$) crystallize in orthorhombically distorted perovskites structure with space group *Pnma*. These orthochromites possesses two structural distortions namely i) rotation (tilts) of the $CrO_6$ octahedra around [010] and [101] which can be quantified by the tilt angles "φ" and "θ" respectively[27] and ii) rare-earth ion ($R^{3+}$) displacement. These two distortions together govern the symmetry lowering from an ideal cubic perovskites type ($Pm\bar{3}m$) to orthorhombic type (*Pnma*) and activate zone-centre Raman modes.[28] Group theory suggests 24 Raman active modes for orthorhombic *Pnma* with four formula units in the unit cell:[29]

$$7A_g+5B_{1g}+7B_{2g}+5B_{3g} \qquad (1)$$

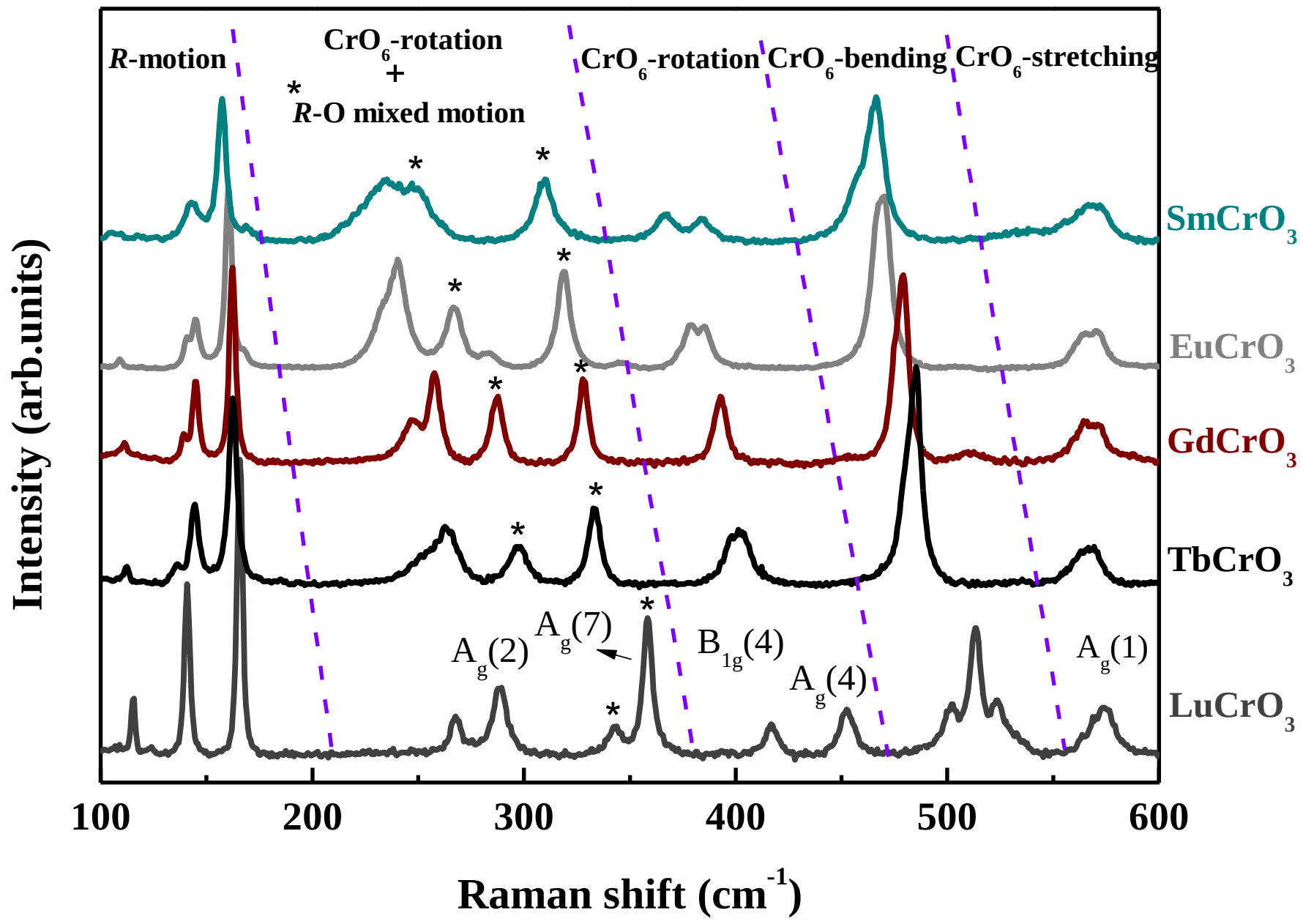

**Figure 1:** (color online) Unpolarized Raman spectra of $RCrO_3$ ($R$= Lu, Tb, Gd, Eu, Sm) recorded at 300 K.

Unpolarized Raman spectra of $R$CrO$_3$ ($R$= Sm, Eu, Gd, Tb and Lu) is shown in Figure 1. Though group theory predicts 24 Raman active modes, we observe only around 11-15 modes. It is possible that certain modes may be too weak to be observed or may be at very low wavenumbers, below our experimental cut off. The detailed analysis of the mode assignment exists elsewhere;[14,30] for the sake of continuity we provide here a discussion on the certain important modes used in our discussion as seen in Figure 1. The sharp and intense modes below 200 cm$^{-1}$ are related to $R$-ion vibrations and are expected to be less affected by the orthorhombic distortion. There are two pairs of CrO$_6$ octahedral rotational modes which are very sensitive to the orthorhombic distortion. Among them, A$_g$(2) and A$_g$(4)  are considered as soft modes and the frequency of these modes vary linearly with tilt angles "φ" ad "θ" respectively.[30]  In Ref. [30], it is mentioned that as the octahedral rotation angle (or the tilt angles "φ" ad "θ") increases, the frequency of the soft modes A$_g$(2) and A$_g$(4) also increases. This provides the measure of deviation from the cubic structure in orthorhombic RCrO$_3$. Hence in cubic structure the frequency of these modes would tend to zero.  The modes which are sandwiched between these pairs of rotational modes (between 250 to 400 cm$^{-1}$) are related to rare-earth and oxygen mixed vibrations in $R$O$_{12}$ polyhedra. Modes related to bending and stretching of Cr-O bonds within the octahedra appear at the high frequency end of the spectra (see Figure 1). Particularly, anti-stretching vibration of the Cr-O bonds inside octahedra (A$_g$(1)) is important as its frequency is very sensitive to the changes in Cr-O bond lengths.

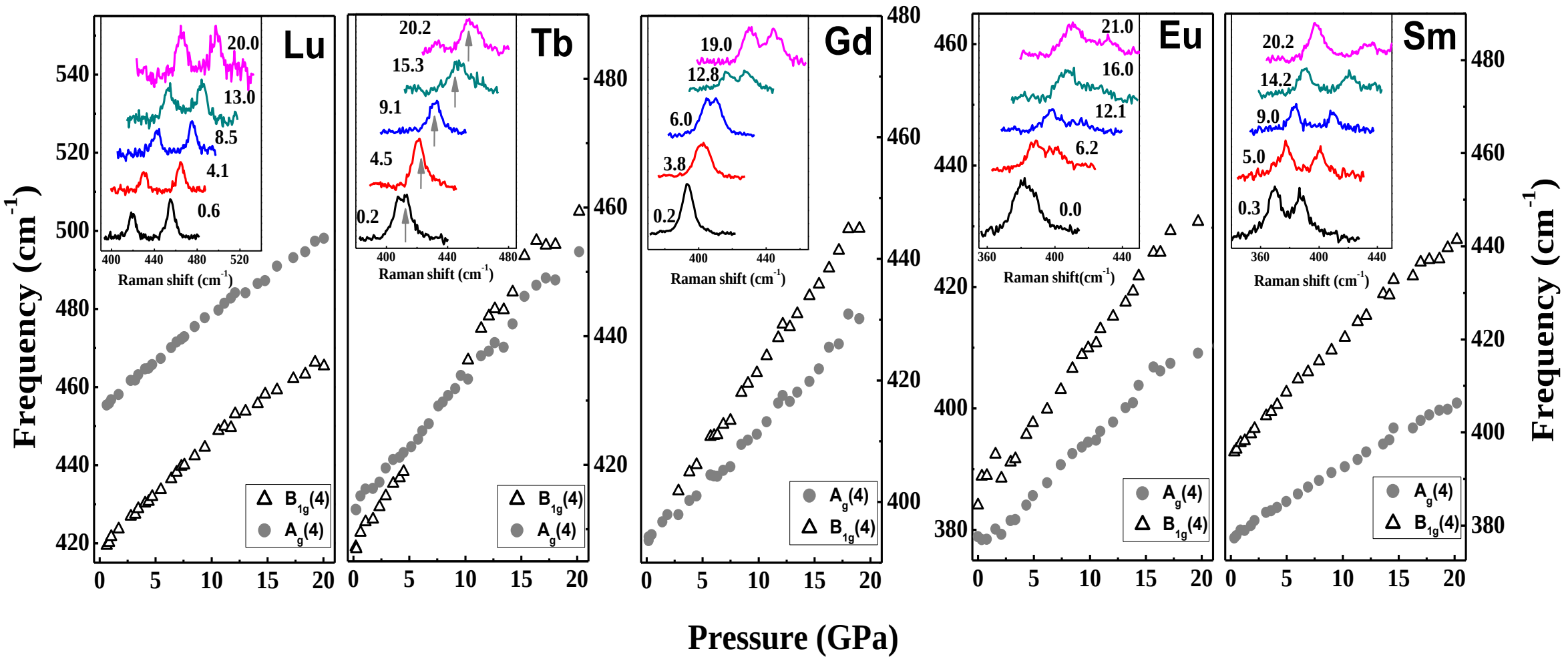

**Figure 2:** (color online) Pressure dependence of rotational modes (A$_g$(4), B$_{1g}$(4)) frequencies in $R$CrO$_3$ ($R$=Lu, Tb, Gd, Eu, Sm). Insets show the mode profiles at different pressures.

Pressure dependence of the rotational modes $A_g(4)$ and $B_{1g}(4)$ are plotted in Figure 2. It is clear that the frequencies of these modes increase with increase in pressure. In particular, the behaviour of $A_g(4)$ mode frequency with pressure can be related to the degree of distortion as this mode frequency varies linearly with the tilt angle ‘θ’. The increase of this mode frequency with pressure indicates that the octahedral distortion is increasing with pressure. The $B_{1g}(4)$ mode can be used for the normalization of the pressure dependence to reveal the behaviour of soft mode ($A_g(4)$) with pressure. It is interesting to see that the separation between these modes is varying in different rare-earths (see Figure 2). The separation remains almost constant in *Lu* up to 20 GPa. These modes merge at 5 GPa in *Tb* and a cross-over is seen beyond 10 GPa. The single mode in *Gd* splits into two and the separation of these two modes increases with pressure. Similarly, in the case of Eu and Sm also the separation increases with pressure. To understand this, we have calculated the rate of change in $A_g(4)$ frequency with pressure ($\Delta A_g(4)$) for different $R$CrO$_3$. We found that $\Delta A_g(4)$ decreases with ionic radii rare-earth cation: $(2.21)_{Lu} > (2.04)_{Tb} > (1.93)_{Gd} > (1.81)_{Eu} > (1.45)_{Sm}$ cm$^{-1}$/GPa. Similar values were obtained for the case of rotational mode $A_g(2)$ (not shown here) which varies linearly with tilt angle “φ”. The trend in $\Delta A_g(4)$ with the size of *R*-ion, indicates that the rate of increase of distortion with pressure is reducing as we go to larger *R*-ions.

In order to understand the role of *R*-ion size for the observed behaviour of octahedral distortions in $R$CrO$_3$ at higher pressure, it is important to study the pressure induced compressions at $CrO_6$ and $RO_{12}$ sites.[16] We made an attempt to compare the compressibilities of $CrO_6$ and $RO_{12}$ by considering the pressure dependence on the Raman modes $A_g(1)$ and $A_g(7)$. It is known that the compressions at A and B sites in orthorhombic $ABO_3$ are anisotropic due to the presence of longer and shorter bonds.[16,18] frequencies of $A_g(1)$ and $A_g(7)$ modes depend on the average Cr-O and *R*-O bond lengths and would be sufficient for the qualitative comparison of $CrO_6$ and $RO_{12}$ compressibility with pressure. The change in these mode frequencies with pressure ($\omega_p$-$\omega_0$) in $LuCrO_3$ and $SmCrO_3$ are shown in Figure 3. However, the the slopes of the linear fits to $\omega_p$-$\omega_0$ vs P for $A_g(1)$ is: $(4.2)_{Lu} < (5.30)_{Tb} < (5.32)_{Gd} < (5.69)_{Eu} < (5.72)_{Sm}$ cm$^{-1}$/GPa. Similarly, for $A_g(7)$: $(1.39)_{Lu} > (0.79)_{Tb} > (0.62)_{Gd} > (0.48)_{Eu} > (0.31)_{Sm}$ cm$^{-1}$/GPa. The trend in these slope values is related to the trend in the compression of the bonds at the respective sites. It clearly indicates that the compressibility of $CrO_6$ is increasing with increase in *R*-ion radii while the compressibility of $RO_{12}$ is decreasing. The reduction in the compression of $RO_{12}$ from *Lu* to *Sm* is associated with the decrease in the change of mean bond distance <*R*-O> which is accompanied by the net increase in the change of mean <Cr-O> bond distance. This is one of the reasons for the decrease in degree of inter-octahedral tilting as seen from the trend in $A_g(2)$ frequency with pressure.

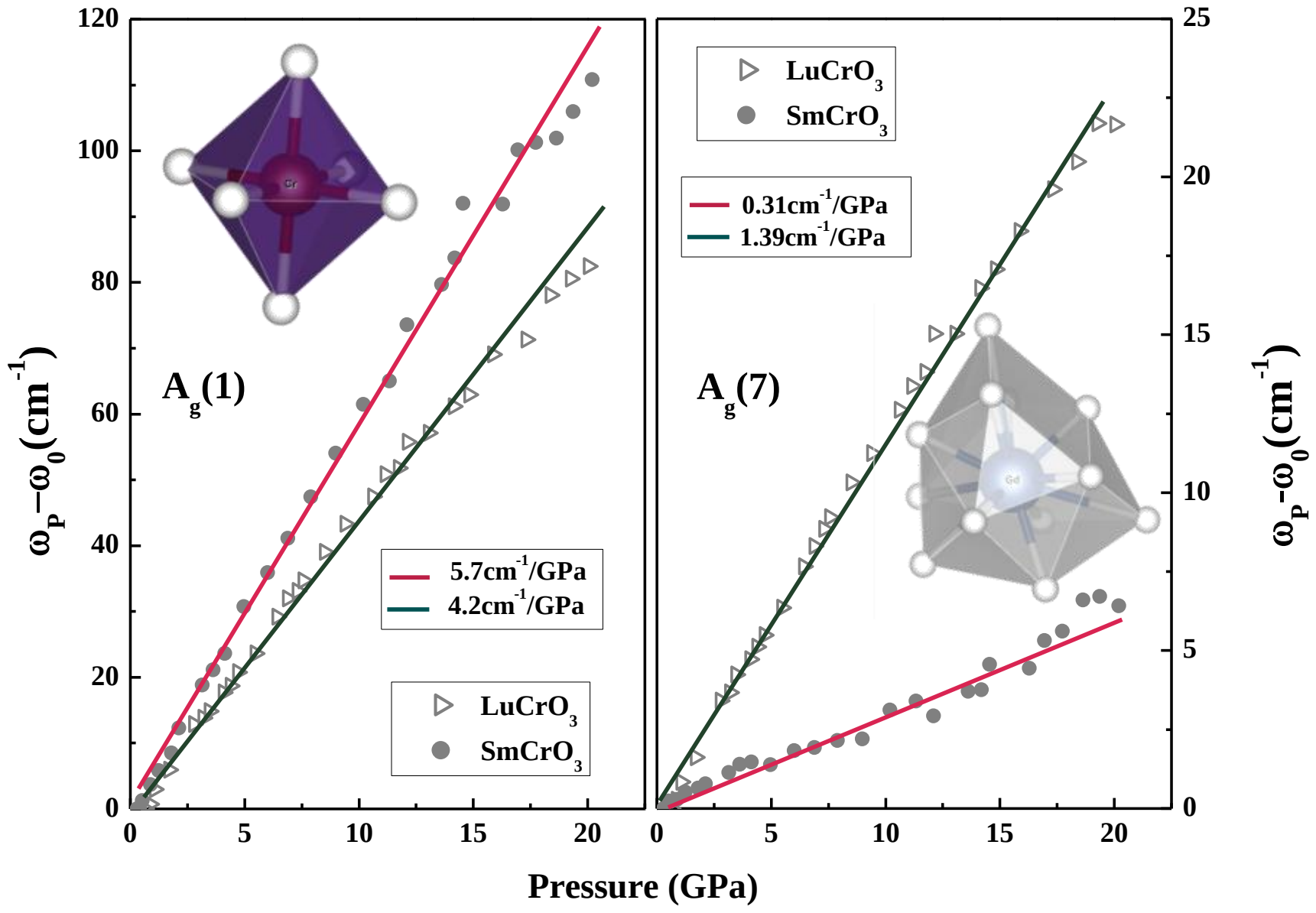

**Figure 3:** (color online) Variation in the Cr-O stretching ($A_g(1)$) and *R*-O vibration ($A_g(7)$) mode frequencies with pressure in $LuCrO_3$ and $SmCrO_3$. The solid lines are ~~the~~ linear fits to the data.

*3.2. X-ray diffraction studies:*

In order to give further insights into the pressure dependence of the octahedral distortion in $R$$CrO_3$, we have done x-ray diffraction studies on $GdCrO_3$, $EuCrO_3$ and $SmCrO_3$. It should be noted that due to the limited beam time available to us at synchrotron, we have carried out the present studies on selected *R*-ion systems. Figure 4 presents the Le Bail profile fits of the corresponding diffraction patterns at ambient as well as at higher pressures. The close agreement of the structural parameters at ambient conditions obtained from the fit with the earlier reports[26] suggests that the sample is of the desired phase.

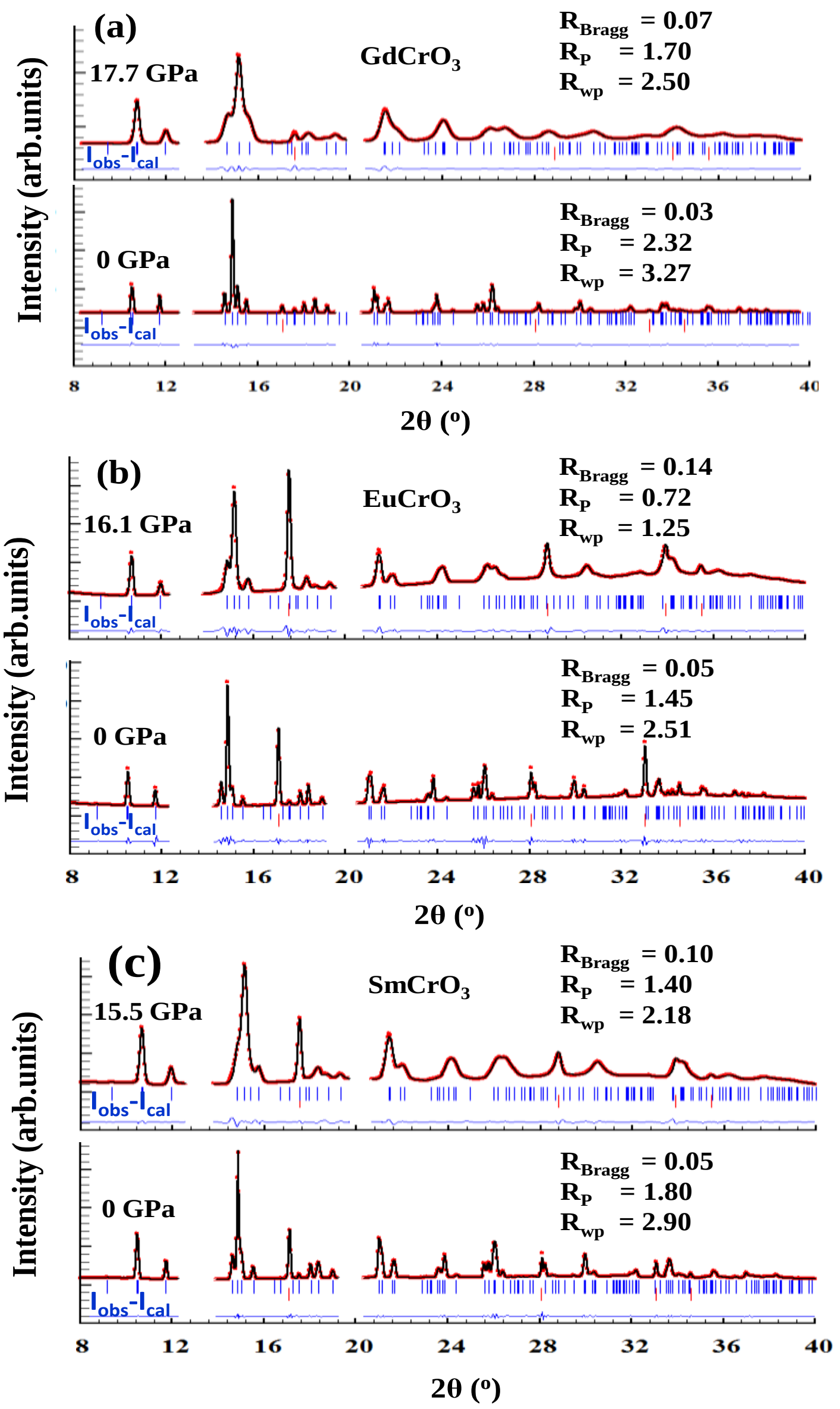


**Figure 4:** (color online) Synchrotron X-ray powder diffraction pattern of (a) $GdCrO_3$, (b) EuCrO3 c) SmCrO3 at different pressures. Red dots represent the experimental data and black solid line is the calculated pattern from Le Bail fit. The blue lines are the reflections from the sample and red ones are from the gold.

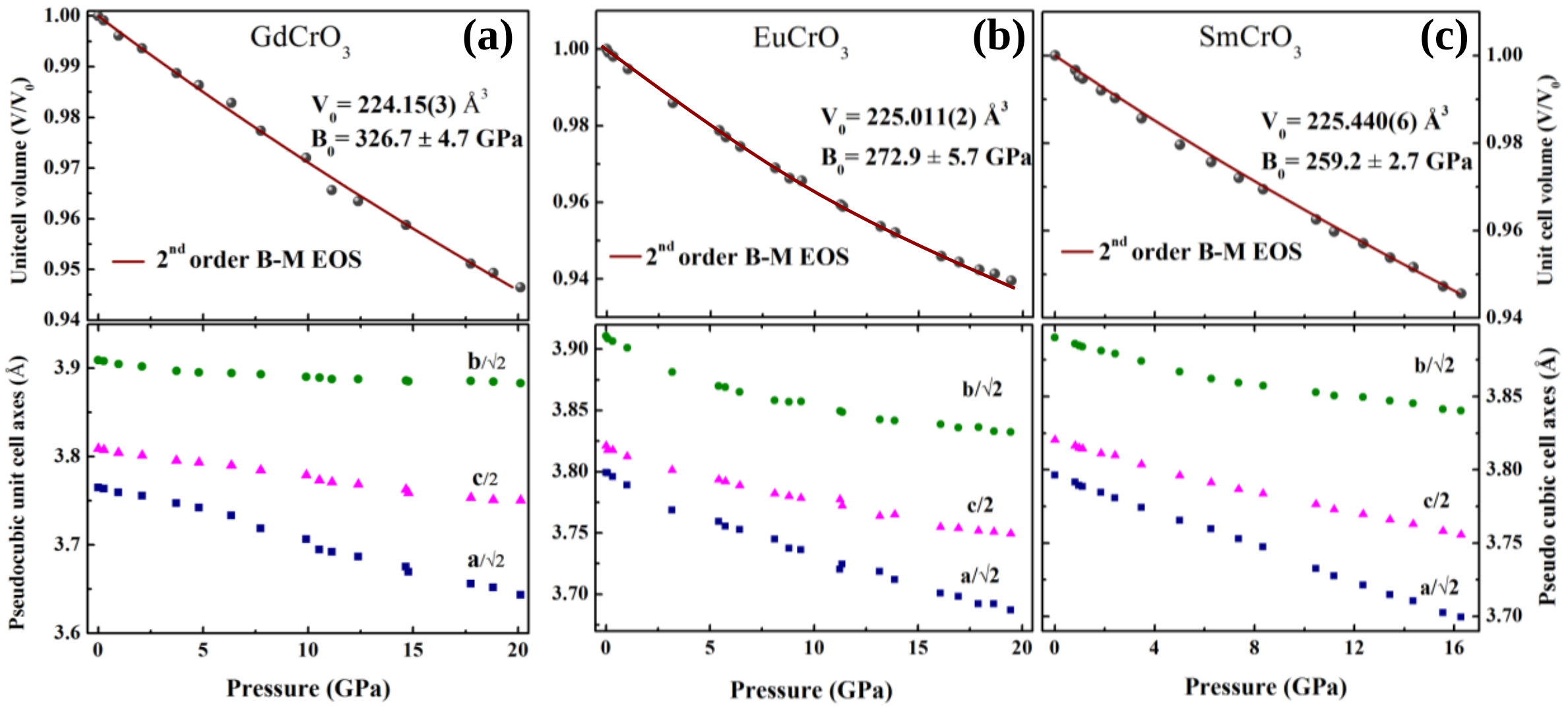


**Figure 5**: (color online) (top panels) pressure dependence of the unit-cell volume of $GdCrO_3$, $EuCrO_3$, $SmCrO_3$ perovskites below 20 GPa. Solid line is the fit of the EOS to the data. (bottom panels) corresponding pseudo cubic unit cell parameters.

The pressure dependence of the unit cell volume of $R$CrO$_3$ is plotted in Fig 5 (top panels). There is no sign of a phase transition within the pressure range of our experiments. The unit cell volume data has been fitted with a 2$^{nd}$ order Birch-Murnaghan equation of state (EOS) given below: [31]

$$P = \frac{3B_0}{2}\left[\left(\frac{V_0}{V}\right)^{\frac{7}{3}} - \left(\frac{V_0}{V}\right)^{\frac{5}{3}}\right] \quad (2)$$

As obtained fit parameters for $V_0$ and $B_0$ are shown in each panel in Figure 5. It is seen that the bulk modulus $B_0$ decreases as $R$-ion radii increases from Gd to Sm.

On the basis of the experimental results of anti-ferromagnetic transition metal oxides, Bloch[32] has put forward a phenomenological rule: $\frac{dT_N}{dP} = 3.3T_N\kappa$, where κ is the compressibility. In general, for perovskites structure, the magnetic transition temperature which depends upon the strength of the exchange interactions ($J$) is directly related to the octahedral tilt angles and bond length, and it is not related to its cell volume. Most of the magnetic perovskite oxides obey the Bloch's rule. Hence, using $\kappa = B_0^{-1}$ from our x-ray diffraction work, we have found out $\frac{dT_N}{dP}$ follows a trend, namely: $(1.686)_{Gd} < (2.230)_{Eu} < (2.509)_{Sm}$ K/GPa.

The pressure dependence of the pseudo-cubic cell parameters ($a_p$, $b_p$ and $c_p$) along $a$, $b$, $c$-axes are shown in Figure 5 (bottom panels). It is to be noted that in $GdCrO_3$, $b$-axis is less compressible compared to $a$ and $c$-axes similar to the $YCrO_3$ case.[19] While in the case of $EuCrO_3$ and $SmCrO_3$, there is compression along all the three axes. The direct outcome of this is the increase in the distortion of the perovskites structure in $GdCrO_3$ compared to that of $EuCrO_3$ and $SmCrO_3$.

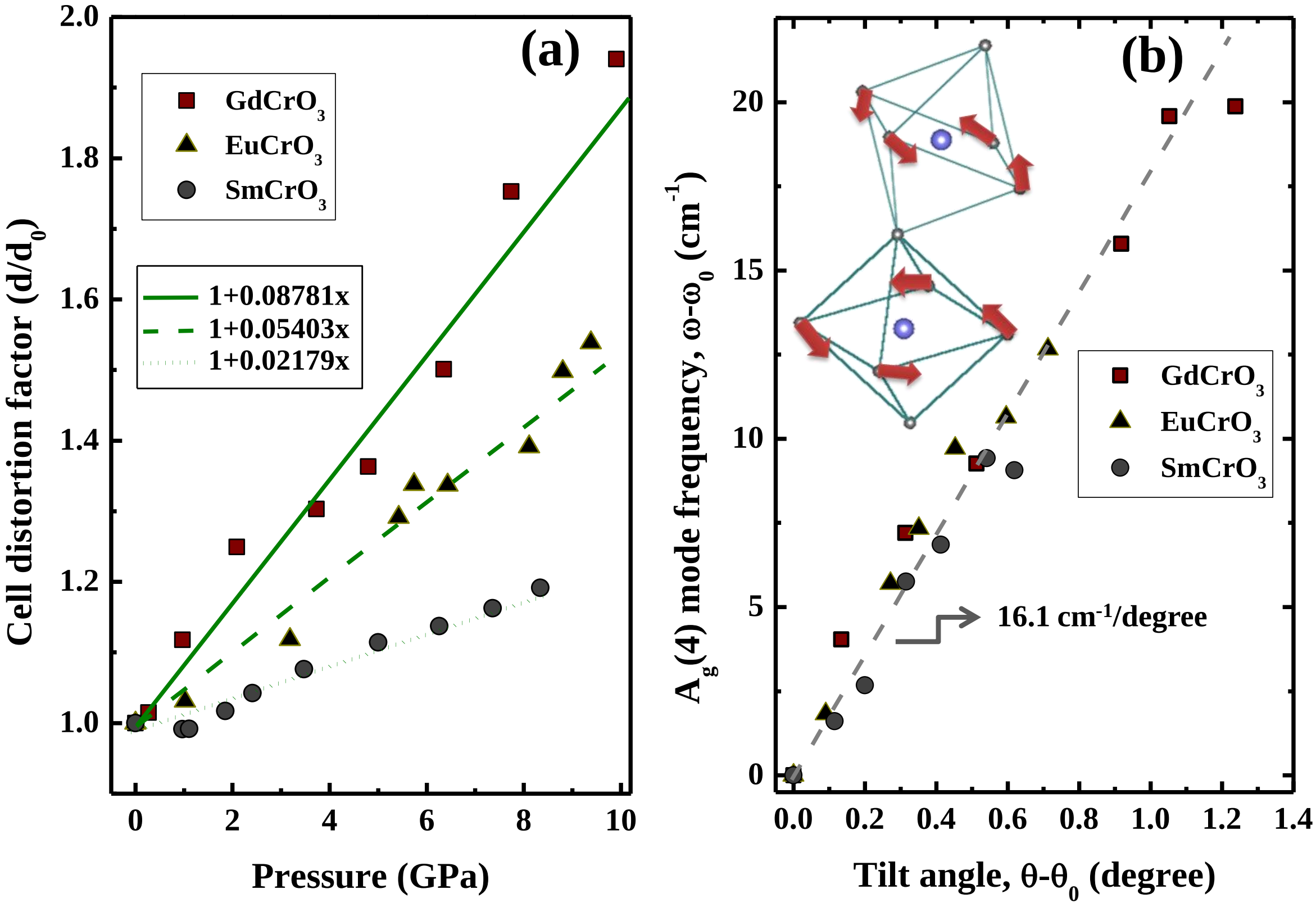


**Figure 6:** (color online) (a) pressure dependence of the cell distortion factor ($d/d_0$) for *R*CrO3 (R-Gd, Eu, Sm). (b) Plot of the change in Ag(4) mode frequency vs the change in tilt angle 'θ'. Dotted line is the linear fit to the data.

The degree of distortion can be estimated from the quantity called cell distortion factor'd' given by: [33]

$$d = \frac{\left[\left(\frac{a}{\sqrt{2}}-a_p\right)^2+\left(\frac{b}{\sqrt{2}}-a_p\right)^2+\left(\frac{c}{2}-a_p\right)^2\right]}{3a_p^2\times10^4} \qquad (3)$$

where, $a_p = \frac{\left[\frac{a}{\sqrt{2}}+\frac{b}{\sqrt{2}}+\frac{c}{2}\right]}{3}$ is the pseudocubic subcell parameters. The cell distortion factor $d/d_0$, increases linearly with pressure (as shown in Figure 6a). It is evident from the slope values

(Δd) of these linear fits that the rate of increase of distortion with pressure is different for different rare-earths. We also have estimated the tilt angles 'θ', 'φ' from the lattice parameters[27] at different pressures and found that the tilt angles increase with pressure indicating the increase in distortion with pressure (not shown here). For instance, change in the soft mode $A_g(4)$ frequency vs tilt angle 'θ' with respect to pressure has been shown in Figure 6b. It can be seen that soft mode frequency scales linearly with tilt angle at a slope of ≈ 16.1 $cm^{-1}$/degree. This slope value is different from the one that is seen in $R$CrO$_3$[30] and $R$MnO$_3$[34] with respect to different *R*-ions suggesting that the effect external pressure on soft mode behaviour is different from that induced by the chemical pressure due to cation substitution. Since pressure applies a far more fine tuned "chemical pressure", this experiment gives a better handle on this value.

This linear relation between the change in soft mode frequency and the tilt angle further prompted us to plot the rate of change in $A_g(4)$ frequency ($\Delta A_g(4)$) with respect to the size of the rare-earth which is shown in Figure 7a. A steep decrease in this value below *Gd* has been noticed. A smooth curve passing through the data points intersects the x-axis around 1.08-1.09 Å, where $\Delta A_g(4)$ =0. Similarly, Δd (obtained from the linear fits shown in Figure 6a) varies linearly with $r_{R^{3+}}$ (as shown in Figure 7b). Interestingly, from the linear fit to the above data we found Δd = 0 for $r_{R^{3+}}$ ≈ 1.09 Å which is exactly where $\Delta A_g(4)$= 0. It means, for the case of $R$CrO$_3$ with $r_{R^{3+}}$ close to the above value, there will not be any pressure induced change in the octahedral tilts. When $r_{R^{3+}}$>1.09 Å, the trend should reverse, namely, the octahedral tilts decrease with pressure, consequently, the perovskites structure will move towards less distorted (more symmetric) phase. The observation of the pressure induced structural transition to rhombohedral symmetry observed in $LaCrO_3$[20] strongly supports our prediction.

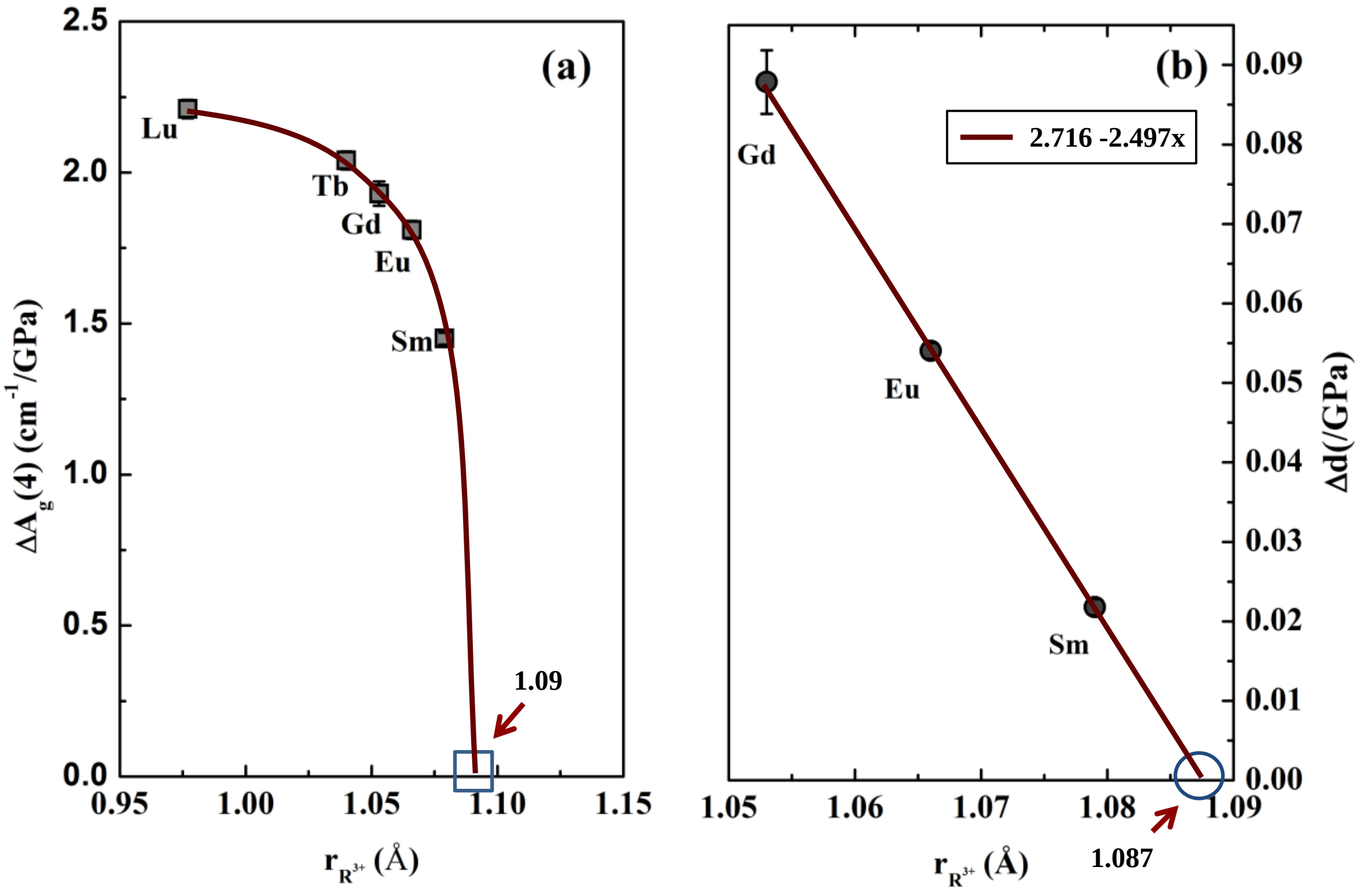


**Figure 7:** (color online) (a) Plot of $\Delta A_g(4)$ against ionic radii of *R*-ion. Red solid line is the smooth curve drawn through the data points. (b) Plot of Δd vs *R*-ion. Red solid line is the linear fit to the data. The points where the curves intersect the x-axis ($\Delta A_g(4)= 0$, $\Delta d= 0$) have been encircled.

It is known that in anti-ferromagnetic $RCrO_3$, the relation between tilt angle and $T_N^{Cr}$ is: [21]

$$T_N^{Cr} \propto \frac{cos^4(\alpha)}{l^7} \tag{4}$$

Here $\alpha = \frac{\theta+2\varphi}{2}$ and $l = \langle Cr - O \rangle$. Then the change in $T_N^{Cr}$ with respect to pressure can be written as

$$\frac{dT_N^{Cr}}{dP} = -T_N \left[\frac{7}{l}\frac{dl}{dP} + \left(2\tan\left(\frac{\theta+2\varphi}{2}\right)\left(\frac{d\theta}{dP} + 2\frac{d\varphi}{dP}\right)\right)\right] \tag{5}$$

Unlike Bloch's rule, Eq.5 shows explicitly the contribution of the tilt angles to the quantity $\frac{dT_N^{Cr}}{dP}$. Due to octahedral compressions under pressure, $\frac{dl}{dP}$ is always a negative quantity. In addition, it is known from Bloch's rule that $\frac{dT_N^{Cr}}{dP}$ is always a positive quantity. As seen from the Raman results here (for instance see Figure 7a) $\frac{d\theta}{dP}$ and $\frac{d\varphi}{dP}$ will possess a non-linear relation with $r_R^{3+}$ below its critical value i.e. $r_R^{3+} \approx 1.09$ Å. Furthermore, it is expected that $\frac{d\theta}{dP}$ and $\frac{d\varphi}{dP}$ will change sign above $r_R^{3+} \approx 1.09$ Å. Thus, one can expect from Eq.5, that these changes in $\frac{d\theta}{dP}$ and $\frac{d\varphi}{dP}$ with $r_R^{3+}$ would reflect in $\frac{dT_N^{Cr}}{dP}$.

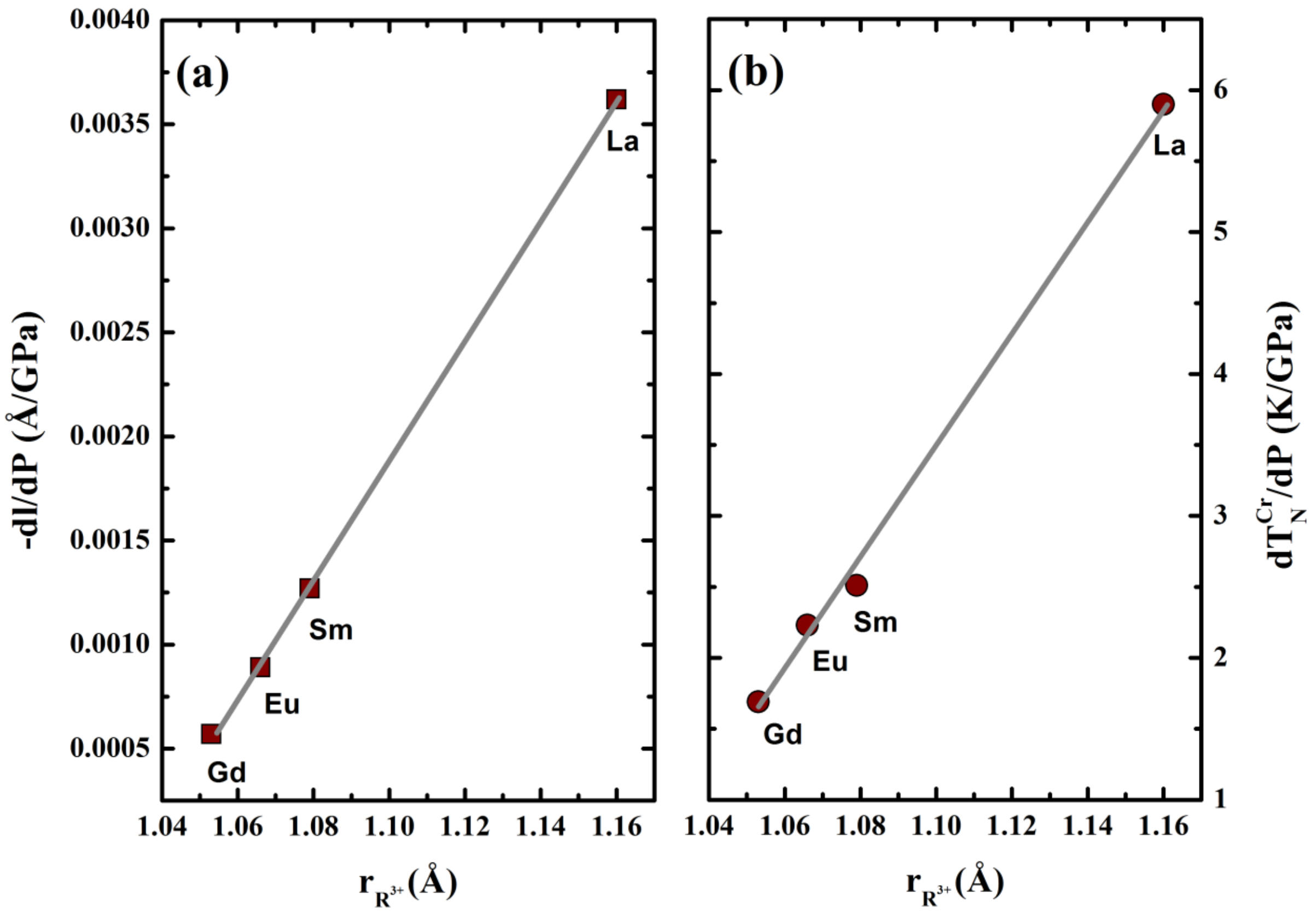

**Figure 8:** Rare-earth ionic radii dependent variation in (a) rate of change in <Cr-O> with pressure. Here <Cr-O> has been estimated form the lattice parameters (Ref.27) (b) rate of change in $T_N^{Cr}$ with pressure calculated using Bloch' rule. The quantities $\frac{dl}{dP}$ and $\frac{dT_N^{Cr}}{dP}$ of $LaCrO_3$ have been calculated from the structural information available in Ref. 20.

However, as shown in Figure 8, $\frac{dT_N^{Cr}}{dP}$ scales linearly with $r_R^{3+}$ similar as $\frac{dl}{dP}$. This clearly indicates that the change in $T_N^{Cr}$ induced by the change in R-ion radii (chemical pressure),[21,35] and in the case of external pressure, compressions at Cr-O plays a dominating role over octahedral tilts in pressure dependent change in $T_N^{Cr}$.

## 4. Concluding remarks:

In conclusion, we have successfully demonstrated the rare-earth (*R*) size dependence of the pressure evolution of octahedral distortions in orthorhombic chromites $RCrO_3$, for the first time, using Raman scattering. Our observations were further supported by the pressure dependent structural data on $RCrO_3$ (R = Gd, Eu, Sm). From the pressure dependence of the octahedral rotational modes as well as the cell distortion factors, we found that the rate of change of the octahedral distortions with pressure reduces with the increase in *R*-ion radii. For a particular *R*-ion radius (~1.09 Å), the rate of change in distortion will be zero and further increase in the *R*-ion radii should lead to reduction in the distortions with pressure. This is the plausible reason for the structural transition observed in the $LaCrO_3$ to a less distorted phase at higher pressures. Pressure induced changes in distortions in perovskites can be related to octahedral tilt angles at room pressure which decrease linearly with increase in *R*-ion radii.[26] When the distortion is high at ambient pressure (as in small *R*-ion case), *R*-ion is less confined in $RO_{12}$ and with pressure $RO_{12}$ compresses significantly leading to a more distorted structure. Conversely, when the distortion is low at room pressure (as in larger *R*-ion), *R*-ion will be strongly confined inside $RO_{12}$ and at higher pressure $RO_{12}$ will be less compressible leading to a less distorted structure. These effects were clearly demonstrated by the pressure dependence of $A_g(1)$ and $A_g(7)$ modes. Furthermore, having the knowledge of the pressure dependent changes in tilt angles for different $r_R^{3+}$, we tried to understand the rare-earth size dependency of $\frac{dT_N^{Cr}}{dP}$ which involves an explicit relation between octahedral bond lengths (<Cr-O>), tilt angles and $T_N$. We found that <Cr-O> plays a dominant role over octahedral tilts in pressure dependent change in $T_N^{Cr}$.

In addition, the spin configuration at low temperatures in $RCrO_3$ strongly depends upon the magnetic $R^{3+}$-$Cr^{3+}$ exchange interactions.[36] The latter is found to be responsible for the magnetoelctric behaviour seen in $RCrO_3$ with magnetic *R*-ion.[13] We would expect that the increase/decrease in the octahedral distortions would enhance/diminish the *R*-Cr interactions. In view of this, it would be interesting to investigate the pressure, temperature effect on the aforementioned properties as a function of *R*-ion radii.

## 5. Acknowledgements:

The synchrotron XRD measurements performed at Elettra synchrotron, Trieste were partially funded by DST, India. Authors are thankful to Mr. B. Rajeswaran for providing the samples,

Mr. Giorgio Bais and Mr. Nicola Demitri for the help in setting up as well as assisting during the experiment at XRD beamline in Elettra. VSB is thankful to Prof. Umesh V. Waghmare and Mr. Nagamalleswararao Dasari for the insightful discussion.